\RequirePackage{lineno}

\documentclass[aps, 10.5pt, superscriptaddress, twocolumn, nofootinbib]{revtex4-2}
\usepackage[UKenglish]{babel}
\usepackage{latexsym}
\usepackage[pdftex, colorlinks=true, linkcolor=red, citecolor=blue, urlcolor=magenta, pdfstartview=FitH]{hyperref}
\usepackage{graphicx}	
\usepackage{amssymb}
\usepackage{amsmath}
\usepackage{xcolor}
\usepackage[normalem]{ulem}

\pdfoutput=1

\usepackage{bm}
\usepackage{verbatim}
\usepackage{amsmath}
\usepackage{amssymb}
\usepackage[utf8]{inputenc}
\usepackage{nicefrac}
\usepackage{braket}
\usepackage{pdfcomment}

\usepackage{pbox}
\usepackage{makecell}
\usepackage{hhline}
\usepackage{multirow}
\usepackage{tabularx}
\usepackage{soul}

\DeclareUnicodeCharacter{2212}{\textendash}

\begin{document}
\title{Polarons shape the interlayer exciton emission of MoSe$_2$/WSe$_2$ heterobilayers}
%

%
\author{P. Soubelet}\email{pedro.soubelet@wsi.tum.de}
\affiliation{Walter Schottky Institut and TUM School of Natural Sciences, Technische Universit\"at M\"unchen, Am Coulombwall 4, 85748 Garching, Germany.}
\author{A. Delhomme}
\affiliation{Walter Schottky Institut and TUM School of Natural Sciences, Technische Universit\"at M\"unchen, Am Coulombwall 4, 85748 Garching, Germany.}
\author{A. V. Stier}
\affiliation{Walter Schottky Institut and TUM School of Natural Sciences, Technische Universit\"at M\"unchen, Am Coulombwall 4, 85748 Garching, Germany.}
\author{J. J. Finley}
\affiliation{Walter Schottky Institut and TUM School of Natural Sciences, Technische Universit\"at M\"unchen, Am Coulombwall 4, 85748 Garching, Germany.}
%
%
\date{\today}
%
%

\begin{abstract}
We present time-resolved and CW optical spectroscopy studies of interlayer excitons (IXs) in 2$H-$ and 3$R-$stacked MoSe$_2$/WSe$_2$ heterobilayers and obtain evidence for the strong participation of hot phonons in the underlying photo-physics. Photoluminescence excitation spectroscopy reveals that excess energy associated with optical excitation of \textit{intra}-layer excitons and relaxation to IXs affects the overall IX-PL lineshape, while the spectrally narrow emission lines conventionally associated with moiré IXs are unaffected. A striking uniform line-spacing of the sharp emission lines is observed together with temperature and excitation level dependent spectra suggesting an entirely new picture that photo-generated phonons lead to phonon-replicas shaping the IX-emission. Excitation power and time resolved data indicate that these features are polaronic in nature. Our experimental findings modify our current understanding of the photophysics of IXs beyond current interpretations based on moiré-trapped IXs.    
\end{abstract}

%
%
\maketitle
%
%


Electron-phonon (e-ph) interactions in bulk-semiconductors play a major role in charge and exciton dynamics where they dominate carrier transport and optoelectronic properties~\cite{sio2023polarons, emin2013polarons, alexandrov2010advances}. In particular, optical phonons most strongly couple to the lattice polarization, allowing the formation of polarons, quasiparticles mediated by the e-ph interaction, and transforming the spectral form of the optical susceptibility.~\cite{sio2023polarons, emin2013polarons, alexandrov2010advances}. For example, e-ph interactions modulate the line shape of interband emission spectra via the emergence of phonon sidebands~\cite{feldtmann2009phonon, shan2005nature, reynolds1995demonstration, grein1990effects, feldtmann2010theoretical}. These effects are expected to be particularly strong in two-dimensional (2D) semiconductors such as monolayer (ML) semiconducting transition metal dichalcogenides (TMDs). These materials feature a direct bandgap in the visible range and efficient light-matter interaction leading to strongly bound excitons~\cite{schaibley2016valleytronics, mak2016photonics, wang2012electronics, Stier.2016, Goryca.2019, mak2012control, Mak2012, Splendiani2010, chernikov2014excitons}. Moreover, the strength of e-ph interactions are enhanced due to the reduced dimensionality and weaker dielectric screening~\cite{sio2023polarons}. 
Experimental and theoretical investigations of the impact of polarons in 2D systems are surprisingly scarce~\cite{sio2023polarons, wang2016tailoring, chen2015observation, cancellieri2016polaronic}, although the fingerprints of polarons have been reported in Raman~\cite{jin2020observation, dyksik2024polaron} and ARPES~\cite{kang2018holstein} experiments.

When two TMD monolayers are vertically stacked to form van der Waals heterobilayers (HBs) their excitonic photophysics becomes even richer. Interlayer hybridization establishes an in-plane periodic moiré potential with a periodicity only determined by the twist angle ($\theta$) and the respective lattice parameters~\cite{van2014tailoring, geim2013van, brem2020tunable}. This new potential folds acoustic phonon branches leading to moiré phonons~\cite{li2023review, quan2021phonon, lim2023modulation, lin2018moire} and tailors collective excitations and the optical response of the system by forming topological exciton bands~\cite{wu2017topological, polovnikov2024field}. Moreover, for HBs having type-II band alignment, such as tungsten diselenide (WSe$_2$) and molybdenum diselenide (MoSe$_2$), the lowest energy excitons form via charge transfer between the layers. They are, therefore, spatially indirect interlayer excitons (IXs), with the electrons located at the bottom of the MoSe$_2$ conduction band and the holes at the top of the WSe$_2$ valence band~\cite{latini2017interlayer}. In this case, the periodic moiré potential confines the IXs, establishing TMD HBs as a promising platform for the study of strongly correlated phenomena and quantum optoelectronics applications~\cite{fogler2014high, wu2015theory, xu2020correlated, regan2020mott, tang2020simulation, huang2021correlated, seyler2019signatures, Shimazaki2019, Wang2020b,Wang2022, Tang2020,regan2020mott,Li2021, Campbell2022}. 

Since IX typically lie several hundred millielectronvolts lower in energy than \textit{intra}-layer counterparts, their formation involves the absorption and emission of phonons to dissipate the excess photon energy.~\cite{nagler2017interlayer, schmitt2022formation, wang2021phonon, ovesen2019interlayer, zimmermann2020directional, katzer2023exciton, sood2023bidirectional, bange2023ultrafast}. Hereby, a large out-of-thermal-equilibrium phonon population is produced by near resonant excitation of \textit{intra}-layer excitons~\cite{nagler2017interlayer, schmitt2022formation, wang2021phonon, ovesen2019interlayer, katzer2023exciton, shinokita2021resonant, sood2023bidirectional, li2023coherent}. Moreover, when an IX is formed, it strains the lattice, changing the equilibrium separation between the constituent TMD layers. Electron-phonon interactions then couple the relative motion of the layers, described through acoustic phonon modes, with the IXs, leading to the formation of IX-\textit{polarons}~\cite{iakovlev2022flexural, semina2020interlayer}. In this letter, we provide compelling experimental proof that the narrow IX emission lines in MoSe$_2$/WSe$_2$ HBs, conventionally ascribed to the recombination of individual moiré-trapped IXs, arise from IXs dressed with an integer number ($N$) of non-thermal acoustic phonons.



\section{results and discussion}
\label{results}

\begin{figure}[t!!]
\includegraphics*[keepaspectratio=true, clip=true, angle=0, width=.99\columnwidth, trim={0mm, 61mm, 180mm, 0mm}]{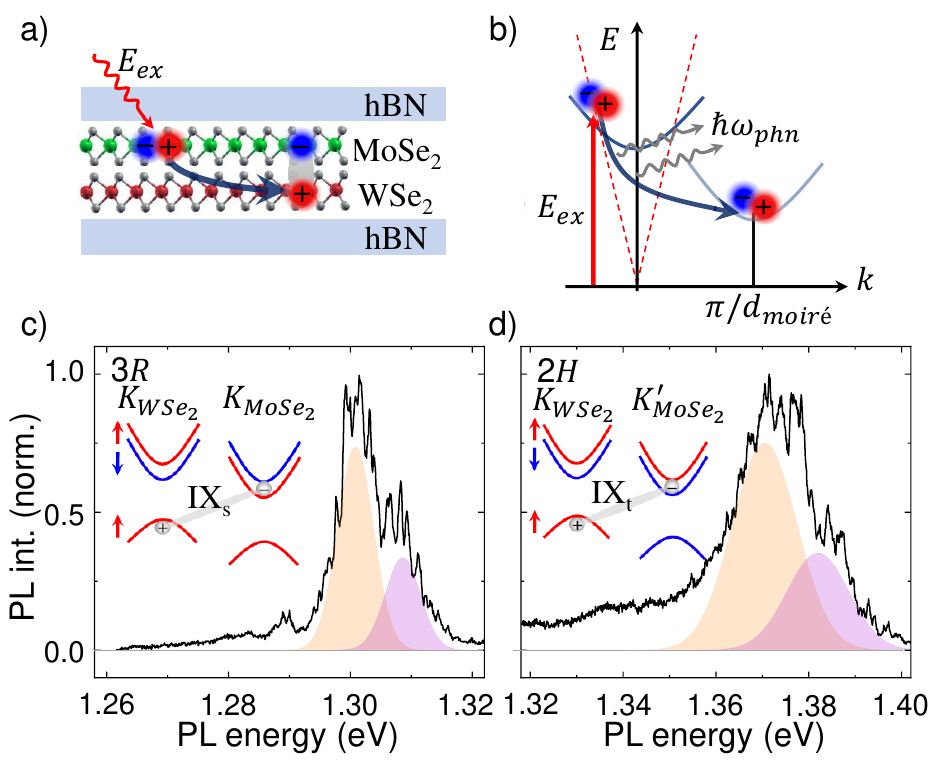}
\caption{\textbf{a} Schematic of the investigated hBN-encapsulated WSe$_2$/MoSe$_2$ HBs. Sketch of the IX formation in \textbf{a)} real space and \textbf{b)} momentum space. An electron-hole pair is excited via the absorption of a photon with energy $E_{ex}$ (red arrow) within the light cone (red dotted lines). During the charge transfer and energy relaxation process, the electron-hole pair strongly interacts with the lattice by absorbing and emitting phonons ($\hbar\omega_{phn}$). \textbf{c)} and \textbf{d)} show the normalized IX emission intensity for the 3$R-$ and 2$H-$ sample, respectively. The general shape of each spectrum can be approximated by two Gaussians (orange and violet shaded regions). Insets in c and d: Schematics of the spin-valley configuration for the 3$R-$ and 2$H-$ sample. The corresponding excitons are indicated.}
\label{figura:1}
\end{figure}

To investigate the physics governing the IX formation and emission, we fabricated HBs via the well-known tear-and-stack method~\cite{kim2016van} to assemble twisted samples with $\theta$ approximately 4.5$^{\circ}$ and 57$^{\circ}$. These twist angles were chosen to allow investigations of near 3$H-$ and 2$R-$stacked optical selection rules, maintaining significant oscillator strength while avoiding lattice reconstruction effects typically arising at small $\theta$s~\cite{woods2014commensurate, weston2020atomic, andersen2021excitons, enaldiev2020stacking, edelberg2020tunable}. A schematic of the hexagonal boron nitride (hBN) encapsulated HBs used in this work is shown in Figure \ref{figura:1}a. Details regarding the fabrication, characterization and optical selection rules of the samples are provided in the Supplemental Material (SM)~\ref{Sample Char}. Although local variations across the samples are present, all key results pertaining to IX-polaron physics were found to be reproducible and independent of the specific location on the samples and twist angle.

The IX formation in real and momentum space is depicted in Figures \ref{figura:1}a and b, respectively. It involves the absorption of a photon of energy $E_{ex}$ within the light cone (red dashed line in Fig.\ref{figura:1}b) and forms an intralayer electron-hole pair in one of the component monolayers. The charge transfer between MLs occurs over fs-timescales through the $\Sigma-$($\Gamma-$) hybridized states of the WSe$_2$(MoSe$_2$) Brillouin zones~\cite{schmitt2022formation, zimmermann2020directional, bange2023ultrafast}. This process generates phonons ($\hbar \omega_{phn}$), leading to the dissipation of hundreds of millielectronvolts from the $intra$-layer exciton to the IX state~\cite{nagler2017interlayer, schmitt2022formation, wang2021phonon, ovesen2019interlayer, katzer2023exciton, shinokita2021resonant, sood2023bidirectional}. In particular, the charge transfer process triggers the emission of the interlayer breathing mode~\cite{li2023coherent, soubelet2019lifetime, jeong2016coherent}. Note that the IX is a spatially and momentum-indirect state, since it is situated at the edge of the mini-Brillouin zone $\sim \pi/d_{moir\acute{e}}$, where $d_{moir\acute{e}}$ is the moiré periodicity (see Fig.\ref{figura:1}b)~\cite{wu2018theory, schmitt2022formation}. Therefore, the interaction with the lattice is intrinsic to the radiative recombination of IXs, as investigated in detail below.

Typical photoluminescence (PL) spectra recorded from the HBs are shown in Figure \ref{figura:1}c and d, consistent with reports in the literature. PL spectroscopy was performed in a He exchange gas cryostat ($T=4.2$\,K) and the optical CW power ($P_{ex}$) from a wavelength-tunable Ti:Sa laser focused to a diffraction-limited spot ($NA=0.82$) was $100\,$nW. The PL spectra for the 3$R$ and 2$H$ samples in figures \ref{figura:1}c and d were obtained by resonant optical pumping of the MoSe$_2$ intralayer exciton at $E_{ex}=1.602\,eV$ and $1.616\,eV$, respectively. The emission is from the singlet(triplet) IX configuration for the 3$R$(2$H$) sample~\cite{shinokita2022valley, brotons2021moire}. For details regarding the exciton complexes and selection rules see SM~\ref{Sel rules}. Typically, IX emission lineshapes are composed of multiple peaks, the details of which depend significantly on the experimental conditions. However, usually, $\sim 10\,$meV broad lobes are attributed to quantized states within the moiré potential (as indicated by the orange and violet shades in Fig.\ref{figura:1}c and d)~\cite{wu2022observation, tan2022signature, wu2018theory, choi2021twist, tran2019evidence}. On top of these broad emission features and specifically at low to medium excitation powers, a number of narrow emission lines (sub meV linewidth) are superimposed. These lines are frequently ascribed to the emission of individual or few IXs confined at local moiré sites that, due to local sample inhomogeneities, lead to emission at varying energy~\cite{brotons2021moire, li2021interlayer, nagler2017interlayer}. We note that at very low excitation power, even narrower emission lines ($\sim 100$ $\mu$eV of linewidth) have been observed exhibiting single-photon emission characteristics~\cite{baek2020highly}. Furthermore, these sharp emission lines have recently been interpreted as interlayer donor-acceptor emission centers.~\cite{cai2023interlayer}. In the following, we show that the MoSe$_2$/WSe$_2$ IX PL, as those presented in Fig.\ref{figura:1}c and d, is shaped by phonon sidebands (orange and violet lobes in Fig.\ref{figura:1}c and d)) and the central dip in the overall spectrum between the broad lobes corresponds to the zero phonon line (ZPL). Moreover, these phonons are coherently generated by the IX during the interlayer charge transfer and energy relaxation process. Therefore, the low(high) energy lobe is formed by processes in which the IX energy is reduced(increased) by the emission(absorption) of acoustic phonons forming exciton-polarons~\cite{semina2020interlayer, iakovlev2022flexural}. The sharp PL lines are phonon replicas from IX-polarons dressed with an integer number of $N-$phonons. While high energy sidebands are frequently suppressed in semiconductor emission~\cite{feldtmann2009phonon, feldtmann2010theoretical}, HBs are a particular case in which the phonons involved in the polaron formation are of low energy, of the order of the thermal energy and, consequently, stimulated processes must be considered.

\begin{figure}[t!!]
\includegraphics*[keepaspectratio=true, clip=true, angle=0, width=1.\columnwidth, trim={-0mm, 0mm, 175mm, 0mm}]{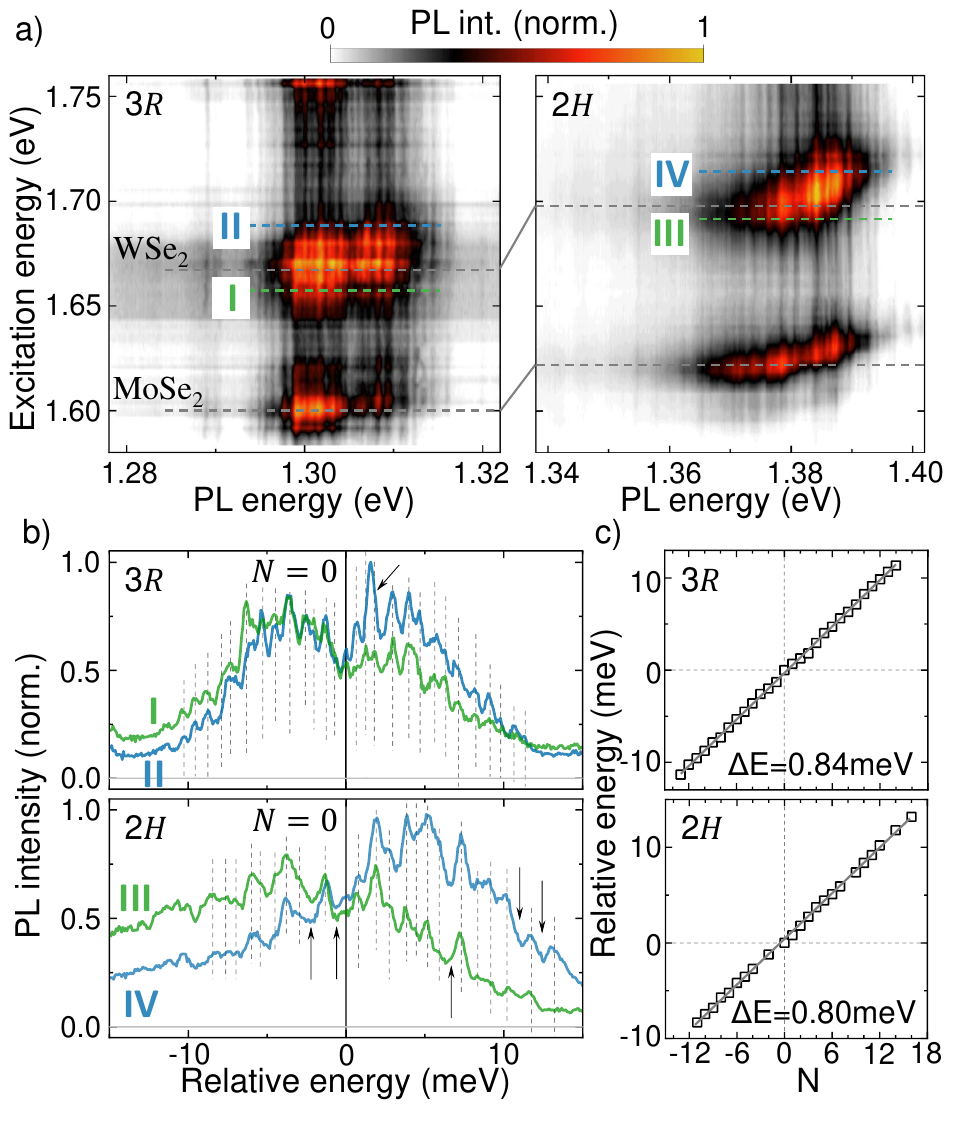}
\caption{\textbf{a)} False color plot of the IX PL spectra for a series of excitation energies $E_{ex}$. \textbf{b)} Selected IX PL spectra of the 3$R-$ (top) and 2$H-$ sample (bottom), plotted on an energy scale relative to the ZPL (peak labeled $N=0$). Spectra in green correspond to the IX emission obtained by exciting the sample just below the intralayer exciton of WSe$_2$ at $E_{ex}=1.653$\,eV(1.687\,eV) for the 3$R-$(2$H-$) sample and the blue spectra are excited just above the resonance, at $E_{ex}=1.685$\,eV(1.720\,eV) for the 3$R-$(2$H-$) sample. Vertical dotted lines mark the position of the sharp PL features. Black arrows mark the position of peaks that were not discernible. \textbf{c)} Peak energy as function of peak number for the 3$R-$ (top) and the 2$H-$ (bottom) sample. The peaks are uniformly spaced with an energy interval of $\sim0.8$\,meV.
 }
\label{figura:2}
\end{figure}

To show that phonons are integral to the IX-PL, we perform PL excitation (PLE) experiments, varying $E_{ex}$ across the MoSe$_2$ and the WSe$_2$ intralayer exciton. As such, we vary the excess energy that needs to be dissipated in the IX formation process. Figure \ref{figura:2}a shows false colour PLE maps for each twisted HB sample. Both samples display an enhancement of the IX emission intensity when the intralayer excitons are pumped resonantly (horizontal dotted lines). The general lineshape of the IX emission depends on $E_{ex}$, highlighted by the selected spectra presented in Fig. \ref{figura:2}b that shows the IX-PL for $E_{ex}$ just above (blue) and below (green) the intralayer exciton resonances, plotted on a relative energy axis with respect to the ZPL. We observe that the intensities of the low and high energy sidebands are affected by $E_{ex}$; pumping above the intralayer exciton resonance enhances the high energy sideband as compared to the low energy sideband. Furthermore, inspecting the PLE map in Fig. \ref{figura:2}a, the high energy sideband blueshifts with increasing $E_{ex}$ as compared to the low energy sideband. We qualitatively explain this observation considering that the phonon population that gives rise to the phonon sidebands depends on the excess energy between $E_{ex}$ and the IX. Strikingly, the narrow IX emission features are completely independent of $E_{ex}$ and, therefore, only depend on the IX recombination process and not on the IX formation process (electron-hole excitation, charge transfer and relaxation processes). Note that the PLE experiments were repeated at higher $P_{ex}$ across multiple spots on each sample, and the observed behavior remained consistent (see SM~\ref{Sample Char II} for details). 

Remarkably, a careful inspection of the sharp polaron peaks across all our experiments (see SM~\ref{Sample Char III}) shows that the overwhelming majority of observable peaks are \textit{uniformly} spaced in energy. This observation cannot be explained with local moiré potential variations\cite{brotons2021moire}, or filling of a single moiré trap with dipolar excitons~\cite{kremser2020discrete}, but rather points towards a specific phonon mode modulating the spectral properties of the IX emission. The uniform energy spacing is plotted versus an integer number $N$ in Figure \ref{figura:2}c, relative to the ZPL, depicted as peak number $0$.  In semiconductors, the vibrational modes that lead to sideband emission are usually lattice modes~\cite{feldtmann2009phonon, feldtmann2010theoretical}. The observed spacing $\sim 0.8\,$meV is an energy that is smaller than the interlayer breathing mode energy of MoSe$_2$/WSe$_2$ HBs~\cite{li2023coherent, lim2023modulation} suggesting that the phonon is located at the $ZA$ acoustic branch~\cite{lin2018moire, quan2021phonon}. In general, phonons involved in the IX emission are located at the edge of the mini-Brillouin zone to absorb the momentum mismatch of the IX relative to the light cone (see Fig.\ref{figura:1}b). Additionally, as the folded phonon dispersion flattens towards the edge of the mini-Brillouin zone, it displays a maximum in the phonon density of states, enabling efficient e-ph coupling. However, the size of the mini-Brillouin zone depends on $\theta$ and, therefore, our samples stacked at 4.5$^{\circ}$ and 57$^{\circ}$ should display different energy spacing, which we do not observe. A second option for the involved vibrational modes are local and quasi-local modes~\cite{brout1962suggested, zaitsev2000vibronic}. These modes are associated with defects and might imply that the presence of the moiré localized IX strongly distorts the lattice and decouples the involved phonons from the general lattice dispersion. In such a case, the polarization of the lattice and therefore the phonon energy may weakly depend on $\theta$, but should have a dependence on the IX density, i.e. on $P_{ex}$. As we show below, such a power dependence was not observed. The precise identification of the phonon mode is out of the scope of this paper in which we limit the discussion to the polaronic nature of the IX emission.

Electron-phonon coupling strength is usually parameterized via the Huang-Rhys factor\cite{mahan2000physics, langreth1970singularities, de2015resolving}. In SM~\ref{Coupling Factor} we calculated the e-ph coupling obtaining an estimated value of $3.1\pm0.4$ and $1.5\pm0.2$ for the samples stacked in the $3R$ and $2H$ configurations, respectively. 

\begin{figure}[t!!]
\includegraphics*[keepaspectratio=true, clip=true, angle=0, width=1\linewidth, trim={0mm, 57mm, 165mm, 0mm}]{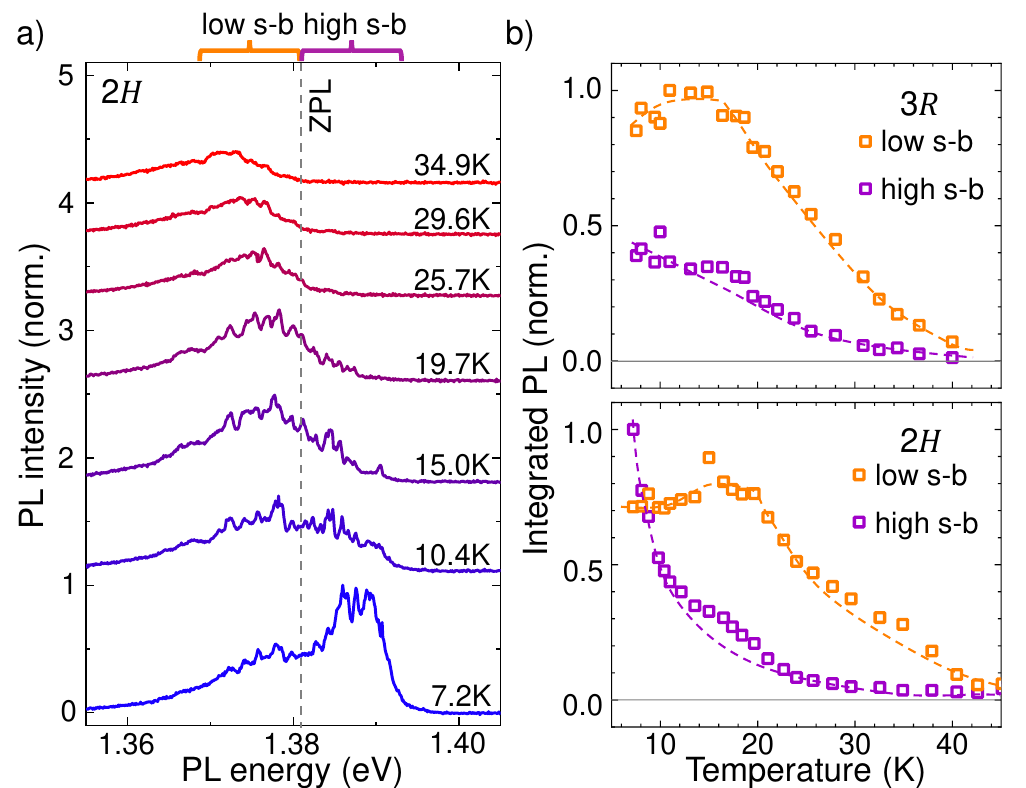}
\caption{\textbf{a)} Temperature dependent PL spectra for the 2$H-$sample. On top, the region of $\pm12\,$meV from the ZPL is marked with orange(violet) to define the low(high) energy sideband. \textbf{b)} PL intensity integrated on the spectral regions defined in a. In orange(violet) is the low(high) energy sideband intensity. Upper panel correspond to the 3$R-$sample and the bottom panel to the 2$H-$sample.}
\label{figura:3}
\end{figure}

To characterize the phonon influence on the IX emission, we performed temperature ($T$) dependent PL spectroscopy. In all experiments, the CW laser power was set to 300\,nW and $E_{exc}=1.96$\,eV. The spectra recorded for the 2$H-$sample are plotted in Figure \ref{figura:3}a (the 3$R-$sample is shown in SM~\ref{Temperature Dep}); all spectra are normalised to the intensity of the $T=7.2\,$K spectrum. The vertical grey dotted line in the figure marks the position of the ZPL. With increasing $T$, the high energy sideband rapidly decreases in intensity, while the low energy sideband initially maintains its intensity and then decreases only above a threshold of $\sim20\,$K. This data is presented in Figure \ref{figura:3}b, where we plot the integrated intensity of each lobe as a function of $T$ for both samples. The narrow emission lines progressively fade out, and cannot be distinguished anymore from the broad background above $30\,K$.  Importantly, we note that they do not shift in energy throughout this $T-$range. Further increasing $T$ above $35\,K$, the emission redshifts and successively decreases in intensity to $\sim 5\%$ of the magnitude at $7\,$K.

The fade-out of the phonon replicas with increasing $T$ is a distinct fingerprint of polaronic states~\cite{feldtmann2009phonon, dyksik2024polaron}. As the phonon population dressing the IX emission is generated through charge transfer and energy relaxation processes, IX and phonons are both out of thermal equilibrium and spatially well-overlapped, specifically enhancing the phonon-IX interactions. The observed $T-$dependence of the high-energy sideband is consistent with stimulated processes. The $N^{th}$ polaron feature on the high-energy sidelobe is an IX dressed with a positive number $N>0$ of phonons. By increasing $T$, the phonon lifetime is reduced monotonically~\cite{duquesne2003ultrasonic, daly2009picosecond} and the IX emission intensity of the high energy sideband reduces accordingly. The low energy lobe is composed of additional spontaneous emission of phonons, a process independent of $T$. Consequently, the low-energy sideband maintains its intensity until a threshold temperature, above which the thermal energy of the crystal affects the radiative lifetime of the IX.  

\begin{figure}[t!!]
\includegraphics*[keepaspectratio=true, clip=true, angle=0, width=1\linewidth, trim={0mm, 56mm, 163mm, 0mm}]{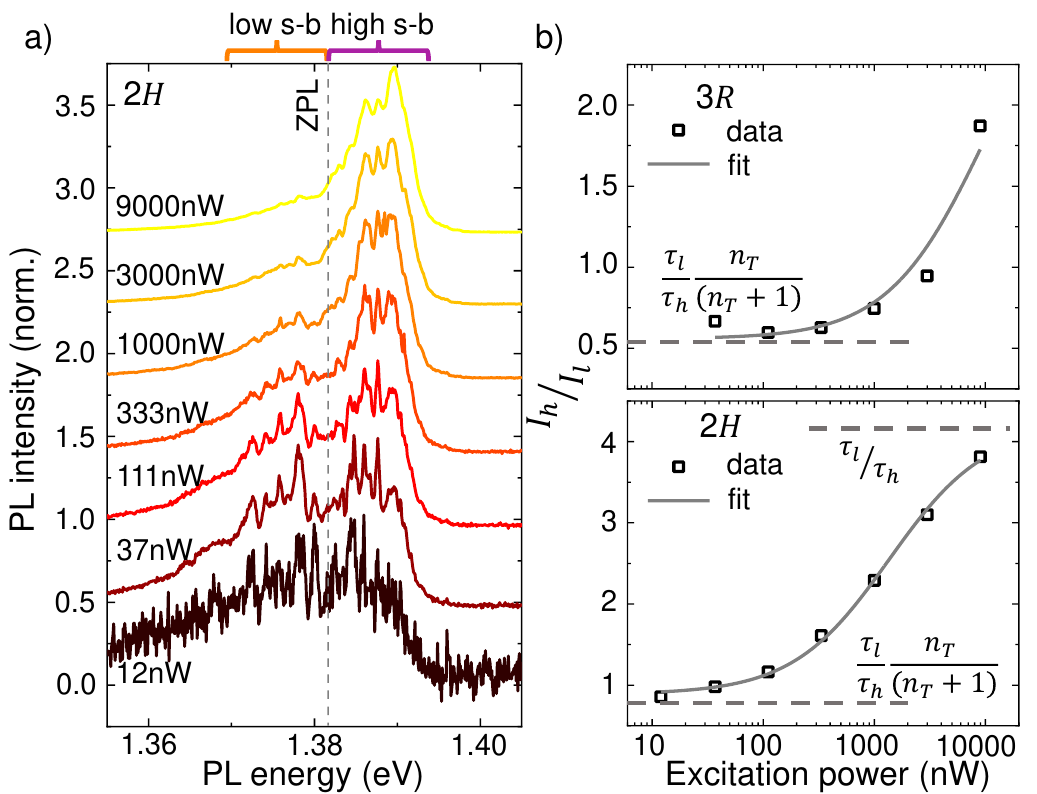}
\caption{\textbf{a)} Power dependent PL spectra for the 2$H-$sample. The region of $\pm12\,$meV around the ZPL is marked with orange(violet) to define the low(high) energy sideband. \textbf{b)} Integrated intensity of the high energy sideband divided by the integrated intensity of the low energy sideband for the 3$R-$ (top) and the 2$H-$ (bottom) sample as a function of excitation power. The transition from a low to a high power regime can be modeled with a rate equation model (solid grey).}
\label{figura:4}
\end{figure}

To support the scenario of non-thermal phonons generated during the IX formation process and the relationship between spontaneous and stimulated emission described above, we perform power-dependent and time-resolved spectroscopy. In Figure \ref{figura:4}a, we present the CW power-dependent PL for the 2$H-$sample, varying $P_{ex}$ from $12\,$nW to $9000\,$nW, at 7\,K with the CW laser at $E_{ex}=1.96$\,eV. The data corresponding to the 3$R-$sample is shown in SM~\ref{Power Dep}. In both cases, the sharp polaron peaks coalesce into a broad emission line with increasing $P_{ex}$. Within the discernible power range, the position of the narrow emission lines does not blueshift, similar to the absence of any $T-$dependent shift observed above. The power dependent PL is accompanied by an increasing dominance of the high energy sidelobe as compared to the low energy sidelobe. This apparent blueshift of the overall IX emission has been interpreted as dipolar IX-IX interactions in the literature ~\cite{brotons2021moire, seyler2019signatures, nagler2017interlayer, li2021optical}. We show in the SM (SM~\ref{Power Dep}) that the sidelobes barely shift in energy, consistent with recent literature taking into account a more comprehensive view on IX interactions beyond simple dipolar repulsive interactions~\cite{steinhoff2023exciton}.

We explain the observed $P_{ex}$ dependence using a phenomenological model, making the following assumptions: i) The phonons contributing to the sidebands are generated during the charge transfer and energy relaxation processes. Consequently, the phonon occupation number is $n = n_T+n_P$, where $n_T$ is the thermal occupation and $n_P=C \cdot P_{ex}$ are the optically generated phonons, proportional to $P_{ex}$ via a proportionality factor $C$. ii) The $N^{th}$ emission line corresponds to $N-$ phonon absorption (for $N>0$) and $N-$ phonon emission (for $N<0$). Note that, strictly speaking, the $N^{th}$ emission line fulfills the following condition: $N=\alpha-\beta$, where $\alpha$($\beta$) corresponds to phonon absorption(emission) processes~\cite{feldtmann2009phonon} and, therefore, it is not possible to fully isolate absorption and emission processes. However, the approximation allows us to determine the relative intensity between the low and high-energy sideband. For instance, the PL intensity of the $N=1$ peak is then composed by the emission from all IXs dressed with one phonon, whose abundance is proportional to $n$. Therefore, its PL intensity is $I_{N=1}\propto n/\tau_{N=1}$, where $\tau_{N=1}$ is the radiative lifetime of this polaron. Analogously, the PL intensity of the $N=-1$ peak is, $I_{N=-1} \propto (n+1)/\tau_{N=-1}$, where $\tau_{N=-1}$ is its radiative lifetime and the term $+1$ is the phonon spontaneous emission probability. The ratio between these quantities becomes
\begin{equation}
\label{eq:1main}
    \frac{I_{N=1}}{I_{N=-1}} = \frac{\tau_{N=-1}}{\tau_{N=1}}\frac{n_T+n_P}{n_T+n_P+1}.
\end{equation}

The expression \ref{eq:1main} leads to two interesting limits that describe the ratio $I_{N=1}/I_{N=-1}$ for low and high $P_{ex}$:
\begin{equation}
\label{eq:2main}
    \lim_{n \to 0} \frac{I_{N=1}}{I_{N=-1}} = \frac{\tau_{N=-1}}{\tau_{N=1}}\frac{n_T}{n_T+1}
\end{equation}
and 
\begin{equation}
\label{eq:3main}
    \lim_{n \to \infty} \frac{I_{N=1}}{I_{N=-1}} = \frac{\tau_{N=-1}}{\tau_{N=1}}.
\end{equation}
At low excitation power, the intensity between low and high energy sideband is proportional to the radiative recombination rates and the thermal occupation. In contrast, at high power, only the radiative recombination rates determine the relative intensities. Further details of the model and, particularly, the extension to $|N|>1$ are discussed in the Supplemental Material~\ref{model}. Without loss of generality and considering the difficulty to isolate individual polaron peaks in the spectra, we analyze the total integrated intensity of the high and low energy sideband $I_h/I_l$ as a function of power. Figure \ref{figura:4}b shows this analysis for the 3$R-$(2$H-$) sample in the top(bottom) panel. In both cases, the experimental data grows from a constant value, in agreement with expression \ref{eq:2main}. By increasing $P_{ex}$, the 2$H-$sample shows a monotonic increase and saturates to a value we identify from equation \ref{eq:3main}. Fitting equation \ref{eq:1main} to the data of the the 2$H$ sample yields $\tau_l/\tau_h=4.2\pm0.1$ and $n_T=0.27$. This thermal population number is consistent with a lattice temperature of  $6\,$K, very close to the experimental conditions. The deduced proportionality constant $C$ is $C=(9\pm1)\times 10^{-4}$\,nW$^{-1}$ and, consequently, $P_{ex}=10\,\mu$W of optical power produces an occupation number of $n_P\sim10$, much higher than the thermal occupation number and consistent with an effective $T$ of $\sim 100\,$K for this particular mode (see SM~\ref{model} for details). In the case of the 3$R-$sample, our data does not reach the limit $n \sim n+1$ and, therefore, the fit to the data assumes $\tau_l/\tau_h=4.2$ as in the 2$H-$sample. In this case, we obtain $n_T=0.23$, a thermal population corresponding to $5.5\,$K and again in good agreement with experimental conditions.

\begin{figure}[t!!]
\includegraphics*[keepaspectratio=true, clip=true, angle=0, width=1.\columnwidth, trim={1mm, 3mm, 150mm, 0mm}]{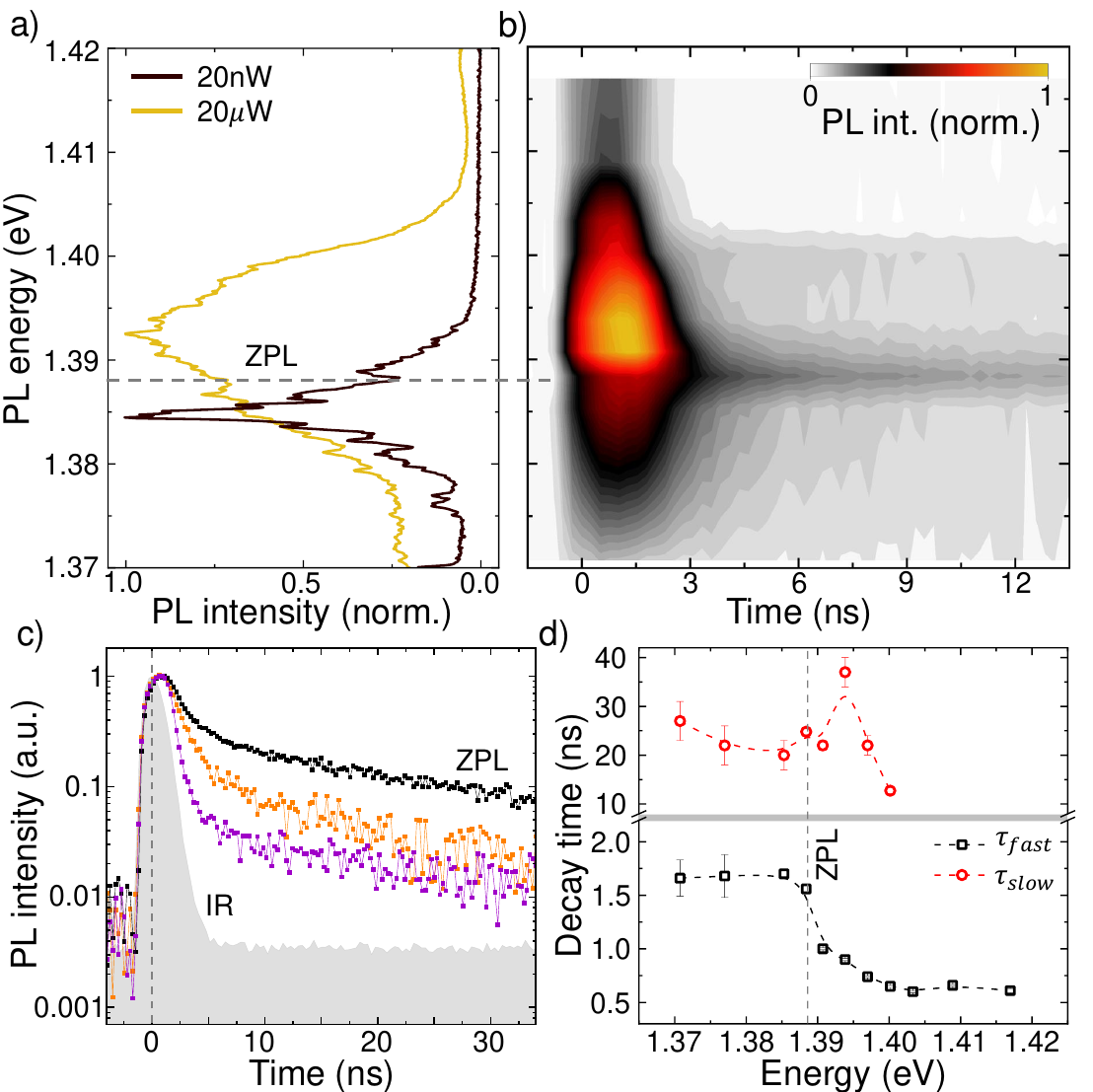}
\caption{\textbf{a)} IX emission of the 2$H-$sample sample using a pulsed laser excitation (pulse duration $1\,ns$, $E_{ex}=632$\,nm) for $P_{ex}= 20\,nW$ and $20\,\mu W$. The ZPL is marked by a grey dotted line. \textbf{b)} False colour map of a time resolved IX emission spectrum for $P_{ex}= 20\,\mu W$. \textbf{c)} PL time traces for $1.385\,eV$ (orange), $1.388\,eV$ (black, ZPL) and  $1.394\,eV$ (violet). A biexponential decay is observed. In grey, the instrument response function (IR). \textbf{d)} Fast (black) and slow (red) decay time obtained from a bi-exponential fit to the data in b).}
\label{figura:5}
\end{figure}

Our phenomenological model predicts that the high-energy sideband might display, on average, a faster radiative decay time than the low-energy sideband. To test this prediction, we performed time and spectrally resolved PL spectroscopy on the 3$R-$sample. The measurements were performed at $T=4\,K$ with a diode laser at $635\,nm$, chopped with a TTL signal to obtain excitation pulses of $1\,ns$ pulse length with a $0.5\,ns$ rise(fall) time. To avoid the re-excitation of the sample before it reaches the equilibrium, the repetition rate was set to $1/200\,$ns. Figure \ref{figura:5}a shows the PL spectra measured with the pulsed laser at an average power of $P_{ex}=20\,$nW and $P_{ex}=20\,\mu$W. As before (see \ref{figura:4}a), there is a clear blueshift of the overall emission, but the ZPL and sharp lines observed in both spectra are independent of excitation power. To time and spectrally resolve the IX emission, we used a monochromator fiber-coupled to a single photon detector resulting in a spectral resolution of $\sim0.75$\,meV. The time-resolved PL shown in the false colour map \ref{figura:5}b was obtained with an average $P_{ex}=20\,\mu$W.

Clearly, the majority of the IX-emission decays in the first few $ns$ after the excitation, while a slower decay extends out into the 10s of $ns$. Strikingly, the ZPL is visible at $\sim 1.388$\,meV, particularly in the long decay trail with higher intensity. We investigate time-dependent PL traces at constant energy above and below the ZPL in Figure \ref{figura:5}c. A clear bi-exponential time dependence is observed, with a fast decay that depends on the PL energy. The time constants for a bi-exponential fit to all-time traces are summarized in Figure \ref{figura:5}d, clearly showing an abrupt change of the fast decay across the ZPL, varying from $\sim1.6\,$ns below ZPL to $\sim 0.6\,$ns above ZPL. Although the latter approaches our instrument response time, $\tau_l/\tau_h\sim2.6$ is in very good agreement with the value obtained from the power-dependent PL fits of our rate equation model discussed above. This internal consistency suggests that the fast decay time of the IX emission is dominated by polaronic decay, where the non-thermal generation of phonons are due to the IX-formation process. Meanwhile, the slower time constant gradually changes across the spectral range with a peak at $1.394\,$eV, 6\,meV above the ZPL, supporting the fact that IX-emission depends on phonons and revealing a longer intrinsic IX lifetime of $\sim40\,$ns, consistent with literature~\cite{blundo2024localisation, rivera2018interlayer, rivera2015observation, nagler2017giant, miller2017long, rivera2016valley, Hollner2024Magneto}.

In summary, we have studied the photo-emission of two MoSe$_2$/WSe$_2$ HBs, with twist angles of $\theta=4.5^\circ$ and $\theta=57^\circ$. Our results strongly suggest that the narrow lines observed in the luminescence of these HBs are attributable to polaronic sidebands. To validate these assumptions, we performed temperature-dependent PL, excitation power-dependent PL, and spectrally and temporally resolved PL experiments. In all cases, we found compelling evidence supporting the validity of the polaronic picture we proposed. The IX formation involves a complex interplay between the optically excited electron-hole pair and the lattice, resulting in a multiple phonon emission and absorption to relax the excess energy from $E_{ex}$ to the IX energy \cite{nagler2017interlayer, schmitt2022formation, wang2021phonon, ovesen2019interlayer, katzer2023exciton, shinokita2021resonant, sood2023bidirectional}. As result of these processes, IX polarises the HB lattice and open a path to interact with phonon acoustic modes~\cite{iakovlev2022flexural, semina2020interlayer}. Considering the low energy spacing between peaks, we propose that the involved phonon belongs to the ZA phonon branch~\cite{lin2018moire, quan2021phonon}. Moreover, as IXs are momentum indirect with a wavevector mismatch of $\pi/d_{moir\acute{e}}$, we assess that the phonon is located at the edge of the mini-Brillouin zone. Our power dependent PL suggests that by increasing the optical excitation power, the generated phonon distribution is significantly higher than the thermal population. This phonon population dominates the IX dynamics and emission at high $P_{ex}$ and allows the observation of the phonon absorption sideband in the IX emission. Our results point to an enhanced interpretation of IX emission, emphasising the need for a deeper understanding of the moiré potentials influence on the coupled IX and lattice dynamics.

%
%

\section{Acknowledgements}

We gratefully acknowledge the German Science Foundation (DFG) for financial support via the SPP-2244 (DI 2013/5-1, FI 947/7-2, FI 947/7-1 and FA 971/8-1) the clusters of excellence MCQST (EXS-2111) and e-conversion (EXS-2089). We additionally acknowledge M. M. Glazov for the fruitful discussions on this project. P.S. acknowledge the financial support of the DFG through the Walter Benjamin program.

The Authors declare no Competing Financial or Non-Financial Interests.


%
%

\onecolumngrid

\section{Supplemental material}

\subsection{Samples fabrication and characterization}\label{Sample Char}

\subsubsection{Samples Fabrication}\label{Sample Fab}

\begin{figure}[t!!]
\includegraphics*[keepaspectratio=true, clip=true, angle=0, width=.9\columnwidth, trim={0mm, 57mm, 200, 2mm}]{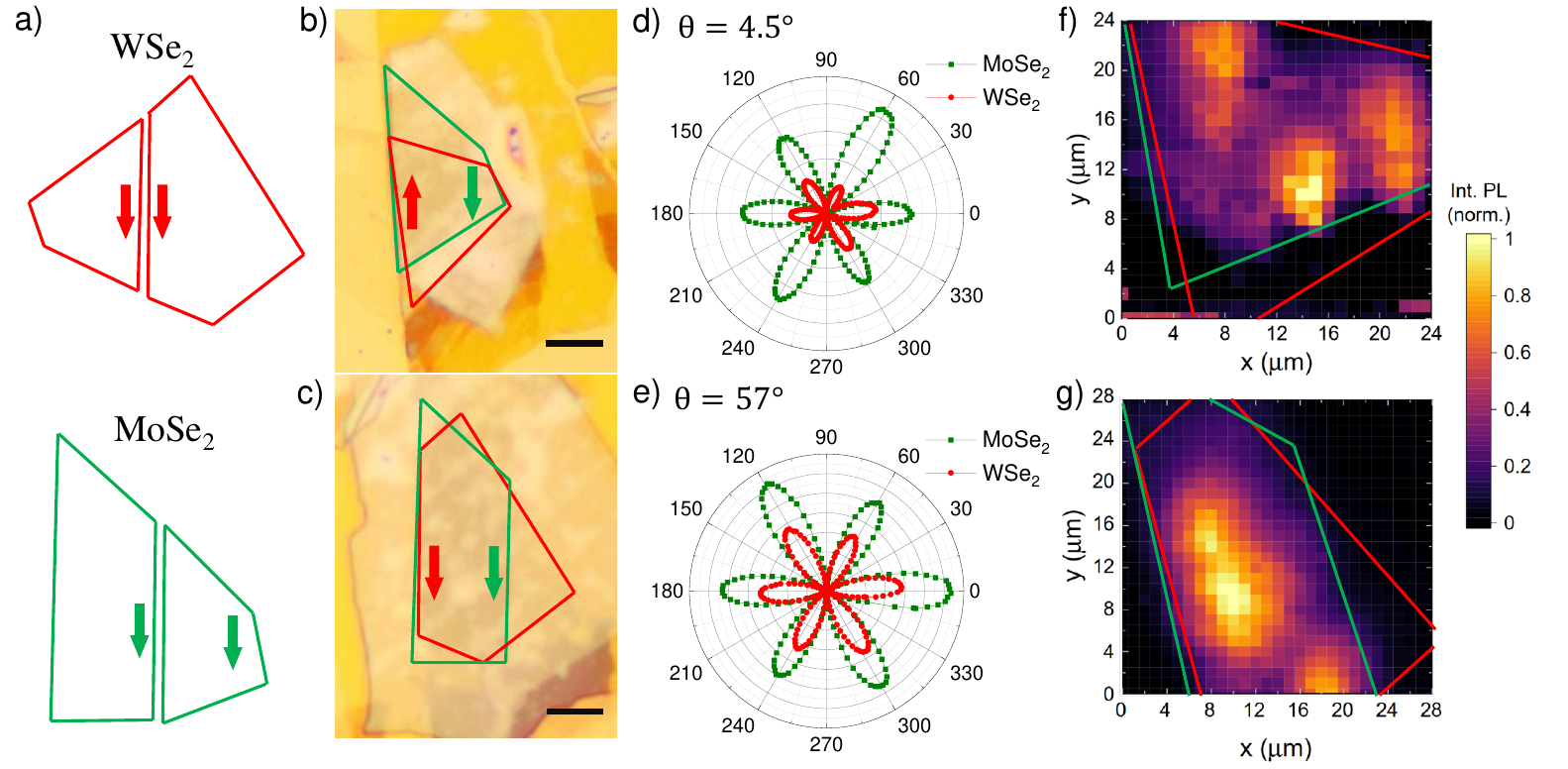}
\caption{\textbf{a)} Sketch of the WSe$_2$ (red) and MoSe$_2$ (green) monolayer and the site in which they were cut during the tear-and-stack process. \textbf{a)} and \textbf{b)} Optical micrographs of the HBs with near 3$R-$ and 2$H-$ stacking, respectively. Green(red) lines indicate the MoSe$_2$(WSe$_2$) monolayer and the scale bar correspond to 8\,$\mu$m. Arrows in \textbf{a}, \textbf{b} and \textbf{c} depict the orientation of the flakes. \textbf{d)} and \textbf{e)} Polar plot of the polarization-resolved SHG intensity in the monolayer region measured for each sample. The relative angle between lobes in \textbf{d} suggest a stacking angle of 4.5$^\circ$ for the sample in \textbf{b} and the plot in \textbf{e} a stacking angle of 57$^\circ$ for the sample in \textbf{c}. \textbf{f)} and \textbf{g)} Integrated PL map in the spectral range of the interlayer exciton emission. for the samples presented in \textbf{b} and \textbf{c}, respectively.}
\label{figura:SM1}
\end{figure}

Monolayer (ML) MoSe$_2$ and WSe$_2$ were obtained from commercial bulk crystals via mechanical exfoliation. The hexagonal crystal structure of a TMD lacks of inversion symmetry and then the crystallographic directions are defined unless a $60^{\circ}$ rotation. Thereby, Both HBs were produced from the same monolayer MoSe$_2$ and WSe$_2$ and each monolayer was divided in two through the tear-and-stack method~\cite{kim2016van}, as sketched in figure \ref{figura:SM1}a. During the fabrication process one of the resulting MLs was rotated by 180$^{\circ}$ as depicted by arrows in figure  \ref{figura:SM1}a and the micrographs in figure \ref{figura:SM1}b and c. During the stacking of each sample, the MLs were aligned to assemble two samples with a known twist angle $\theta$ of approximately 4.5$^{\circ}$ and 57$^{\circ}$. In MoSe$_2$/WSe$_2$ HBs stacked near 0$^{\circ}$(60$^{\circ}$), atoms within each layer try to adjust the stacking landscape forming reconstructed commensurate regions~\cite{woods2014commensurate, weston2020atomic, andersen2021excitons, enaldiev2020stacking, edelberg2020tunable}. Therefore, we used twist angles above 2.5$^{\circ}$ and below 59$^{\circ}$ for 2$H-$ and 3$R-$ stacking, respectively, to avoid reconstruction~\cite{weston2020atomic, andersen2021excitons, enaldiev2020stacking}. The samples were encapsulated between thin hexagonal boron nitride (hBN) flakes using dry transfer techniques based on polycarbonate films, similar to Ref. \onlinecite{castellanos2014deterministic}. 

The substrates used in this work correspond to piezoelectric strain actuators as those used in Refs.\onlinecite{PhysRevB.108.L041404, martin2017strain, iff2019strain, ziss2017comparison, ding2010stretchable} and, therefore, have evaporated gold on the surface. However, it is important to note that we do not present in this work any effect related with strain as well as we do not expect any effect of the gold substrate on the TMDs emission, in agreement with Ref.\onlinecite{paradisanos2021efficient}. 

The stacking angle was confirmed by performing polarized resolved Second Harmonic Generation (SHG) experiments \cite{hsu2014second}. Figure \ref{figura:SM1}d and e, present the polar plot of the polarized resolved SHG from which the stacking angle was determined.

Figure \ref{figura:SM1}f and g present the interlayer exciton integrated emission for the 3$R-$ and the 2$H-$sample, respectively. While the shape of the emission slightly depends on the position in the sample, all measurements presented in this work correspond to regions in which the emission was spectrally sharp and homogeneous. We observed that, in all such regions, the tendencies and features observed and presented are consistent.

\subsubsection{Optical selection rules.}\label{Sel rules}

\begin{figure}[t!!]
\includegraphics*[keepaspectratio=true, clip=true, angle=0, width=1.\columnwidth, trim={-70mm, 10mm, 255, 2mm}]{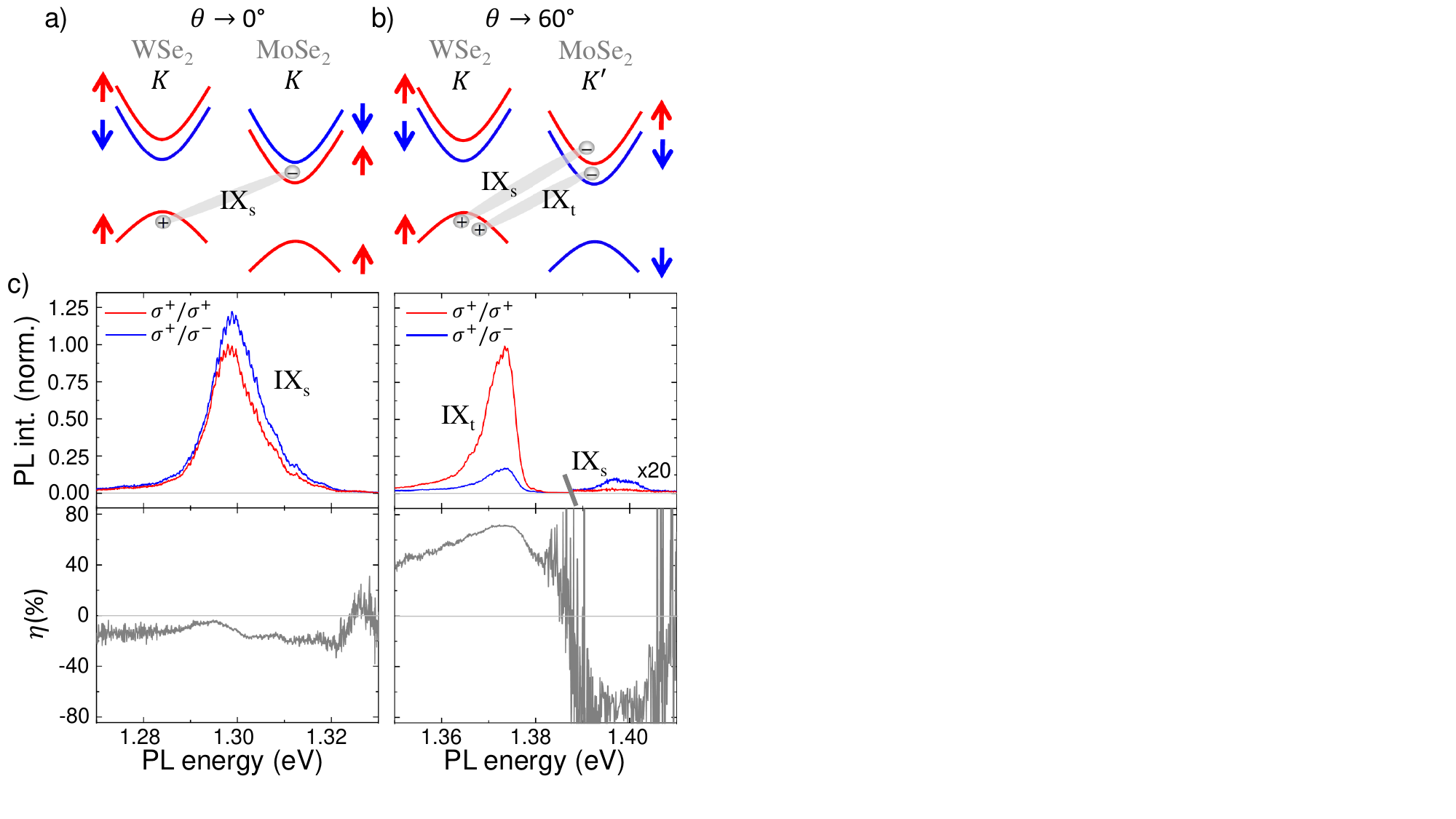}
\caption{\textbf{a)} and \textbf{b)} Schematics of the spin-valley configuration for samples with stacking angle $\theta \rightarrow 0^{\circ}$ and $\theta \rightarrow 60^{\circ}$, respectively, the corresponding exciton complexes are plot over the electronic dispersion of each sample. \textbf{c)} Polarized resolved interlayer exciton PL (top) and circular polarization degree (bottom) for the samples stacked at 4.5$^\circ$ (left) and 57$^\circ$ (right), respectively. The spectra are normalized to the maximum intensity of the co-polarized PL in each case. 
}
\label{figura:SM2}
\end{figure}

In MoSe$_2$/WSe$_2$ HBs, the 3$R-$stacking positions the WSe$_2$ $K$($K'$)-valley at the same point in the k-space as the MoSe$_2$ $K$($K'$)-valley and the 2$H-$stacking aligns the WSe$_2$ $K$($K'$)-valley with the MoSe$_2$ $K'$($K$)-valley. However, the optical selection rules are governed not only by the valley and spin degrees of freedom but also by the Bloch phase factor within the three-fold rotational symmetry $\hat{C}_3$~\cite{yu2018brightened, shinokita2022valley, wang2019giant}. The band alignment of the 3$R$ and 2$H$ HBs are illustrated in the Figure \ref{figura:SM2}a and b, respectively. Additionally, the sketches includes the exciton complexes observed in this work that were also previously reported in the literature, they are a singlet interlayer exciton (IX$_s$) in both kind of stacking and the triplet interlayer exciton (IX$_t$) in the 2$H$ case~\cite{shinokita2022valley, brotons2021moire}. 

Figure \ref{figura:SM2}c shows in the upper panels the circularly co- and cross-polarized photoluminescence (PL) for the 3$R-$ and the 2$H-$sample at left and right panels, respectively. These measurements were performed with a CW 700$\,$nm laser and an excitation power of 40$\,\mu$W. Under the PL spectra, the lower panels of Figure \ref{figura:SM2}c show the circular polarization degree, defined as $\eta=(I^+-I^-)/(I^++I^-)$, were $I^+$($I^-$) is the PL intensity in the co-(cross-)polarized experiments. Although we are not using the conventional 3$R-$ and 2$H-$stacking, the sample stacked at 4.5$^{\circ}$(57$^{\circ}$) exhibits circularly cross-polarized(co-polarized) photoluminescence (PL), consistent with an interlayer emission from moiré sites in samples with 3$R-$(2$H-$)type structures~\cite{yu2018brightened, mahdikhanysarvejahany2022localized, brotons2021moire, shinokita2022valley, yu2017moire, seyler2019signatures}.

\subsection{PLE experiments at different excitation power}\label{Sample Char II}

\begin{figure}[t!!]
\includegraphics*[keepaspectratio=true, clip=true, angle=0, width=1.0\columnwidth, trim={0mm, 50mm, 40, 0mm}]{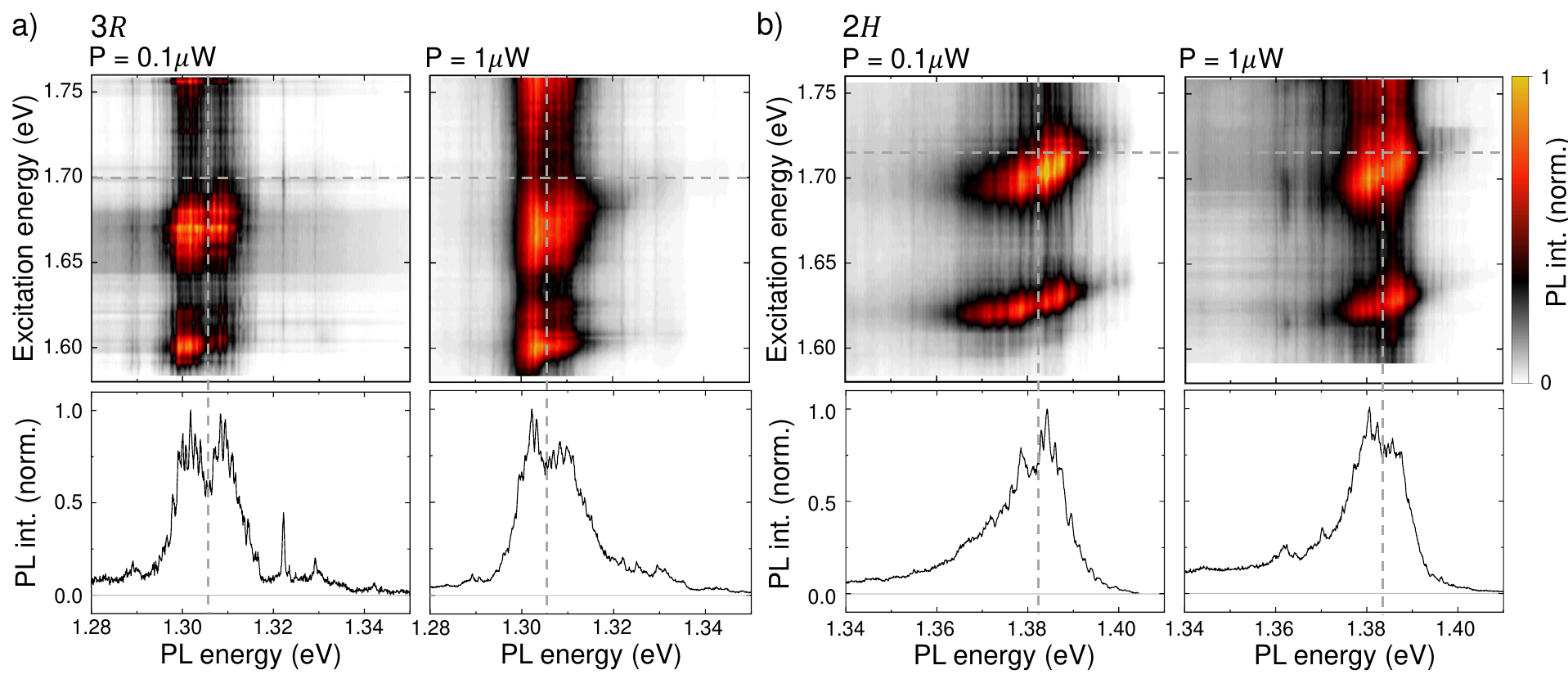}
\caption{\textbf{a)} and \textbf{b)} PLE experiments at different excitation power for the sample stacked in 3$R$ and 2$H$ configuration, respectively. On left and right panels are the experiments performed at $P_{ex}=0.1\,\mu$W and $P_{ex}=1\,\mu$W, respectively. Top panels: False color map of the IX emission as function of $E_{ex}$. The horizontal dotted line marks the selected spectra displayed in the bottom panels. Vertical dotted lines mark the ZPL.
}
\label{figura:SM-3}
\end{figure}

Figure \ref{figura:SM-3} displays the PLE experiments for the 3$R-$ and 2$H-$sample, respectively. Each figure compare the experiments performed at $P_{ex}=1\,\mu$W and $P_{ex}=10\,\mu$W. In this range of energies, while the sharp peaks are clearly distinguished, the observations are independent of $P_{ex}$ and are the same as those discussed in the main text.

\subsection{Characterization of the phonon replicas}

\subsubsection{Additional information regarding the normal distribution of peaks in the PL}\label{Sample Char III}

\begin{figure}[t!!]
\includegraphics*[keepaspectratio=true, clip=true, angle=0, width=1\columnwidth, trim={1mm, 7mm, 18mm, 2mm}]{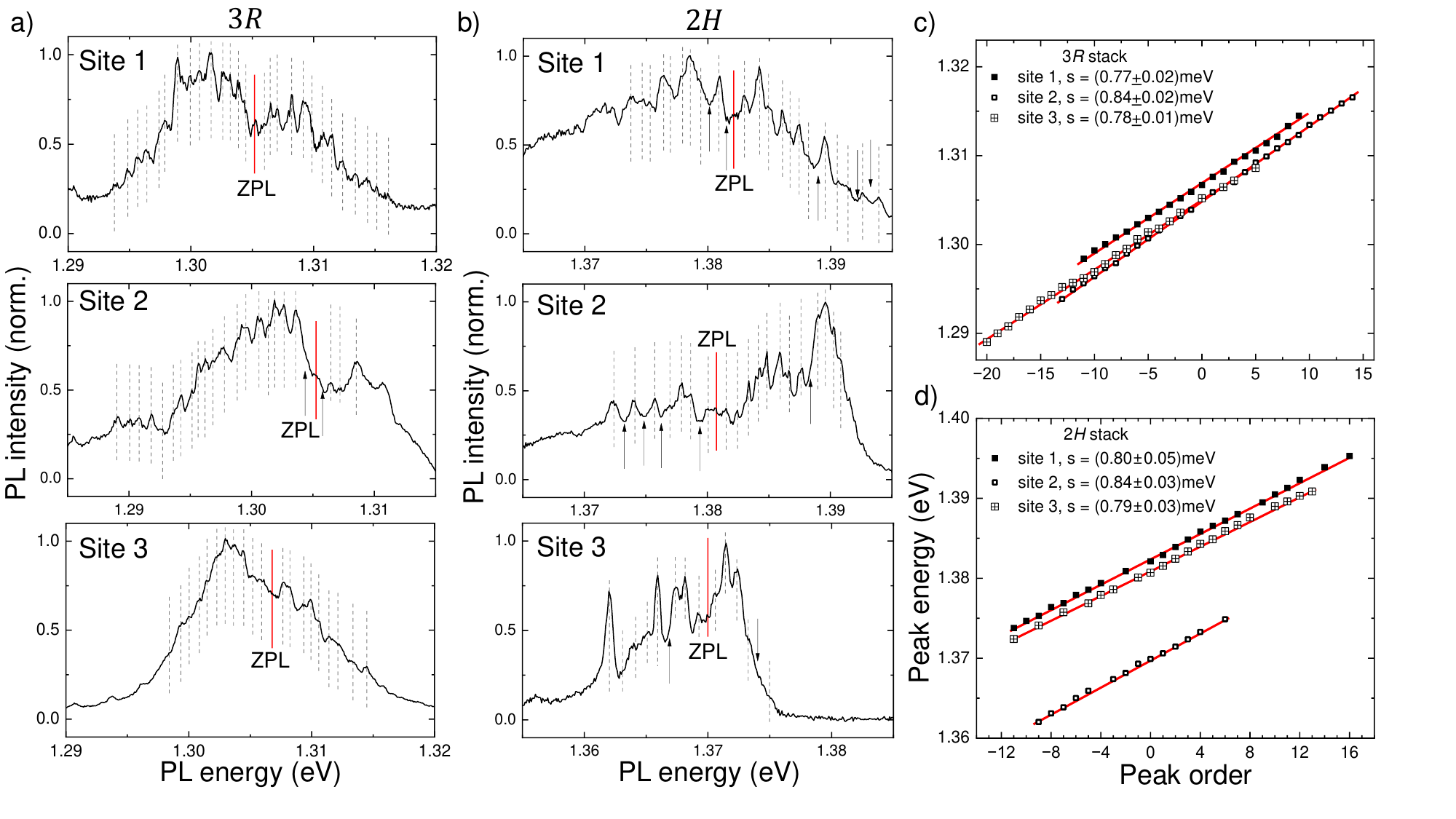}
\caption{\textbf{a)} and \textbf{b)} present the interlayer exciton emission in three different spots on each sample, 3$R-$ and 2$H-$sample, respectively. On each panel, the emmission peaks are marked with dotted vertical lines and the zero order with a red line. \textbf{c)} and \textbf{d)} Peak energy as function of peak order for the spectra in \textbf{a)} and \textbf{b)}, respectively.}
\label{figura:SM-5}
\end{figure}

Figure \ref{figura:SM-5}a and b present, for the 3$R$ and the 2$H$ HBs, respectively, the PL spectra taken at different positions of each sample. While the overall emission lineshape varies along the sample, the narrow lines keep the spacing of $\sim$0.8\,meV. This shows that the phonon mode involved in the emission presents a defined energy that rise from the moiré lattice, averaging many sites. The presence of higher disorder in the lattice is then expressed by the broadening of those lines and not by the peaks separation. For instance, the third site of the 3$R$ sample displays a PL in which the peaks are comparatively much broader and are only distinguishable as shoulders in the overall emission.

It is important to note that all experiments presented in the main text were performed on the first and second spot on each sample and, for this reason, the ZPL was observed in PLE, temperature dependent PL or power dependent PL. For this reason, the ZPL of the third spot on each sample is not properly defined. In those cases the peak marked as ZPL is merely an estimate.

\subsubsection{Additional information regarding electron-phonon coupling factor}\label{Coupling Factor}

The phonon-assisted PL equations can be solved analytically to calculate the steady-state intensity of phonon replicas of arbitrary order \cite{feldtmann2009phonon, feldtmann2010theoretical}. However, these theoretical descriptions are based on some assumptions that do not apply to IXs in HBs. First of all the calculations are based on the fact that, generally, the thermal energy is much lower than that of the phonon modes involved in the electron-phonon coupling. Consequently, only spontaneous emissions are considered in the process, resulting in red-shifted phonon sidebands. In our case, the electron hole couples produce a cascade process, emitting phonons to relax the excess energy from the $intra$-layer exciton energy to the IX state. Therefore, not only the thermal energy is higher than the phonons implied in the e-p coupling, but also the stimulated processes cannot be neglected. 

As we lack of theoretical background including the stimulated processes in the polaron description, we calculate the electron phonon coupling as usual, using only the low energy phonon sideband. Polaron spectra are frequently described through a summation of Lorentzian profiles of the form
\begin{equation}
\label{eq:0main}
    \sum_{N} = \frac{A_N \left(\frac{\Gamma_N}{2}\right)^2}{(\omega-\omega_N)^2+\left(\frac{\Gamma_N}{2}\right)^2},
\end{equation}
where $\omega_N$ is the central frequency of the $N$th peak and $\Gamma_N$ its linewidth. In the strong coupling regime~\cite{mahan2000physics, langreth1970singularities, de2015resolving}, the individual peak intensity $A_N$ is well described by a Poisson distribution function
\begin{equation}
\label{eq:01main}
    A_{N} = A_0\frac{e^{-\alpha} \alpha^N}{N!}
\end{equation}
where $\alpha$ is the Huang-Rhys factor that characterize the strength of the e-ph coupling and $A_0$ is the peak intensity of the ZPL. By fitting expression \ref{eq:0main} to the spectra from Fig.\ref{figura:2}b it is possible to obtain $A_N$ and calculate $\alpha$ through expression \ref{eq:01main}.

Figure \ref{figura:SM-5}a and b presents, for the 3$R-$ and the 2$H-$sample, respectively, the fitting of the spectra with multiple Lorentzian curves. Note that in our case, it presents some difficulties because the distance between phonon replicas is similar to their linewidth. The peaks used for the calculation are marked with a grey rectangular shape. Figure \ref{figura:SM-5}c and d display the extracted peak intensities that, although noisy, can be fitted with a Poisson distribution curve resulting in a Huang-Rhys factor of $3.1\pm0.4$ and $1.5\pm0.2$ for the 3$R-$ and the 2$H-$sample, respectively. Note that in expression \ref{eq:01main} $\alpha$ is the e-ph coupling in three dimensions, we have scaled it by a factor $3\pi/4$ for 2D systems~\cite{peeters1987scaling, jin2020observation}.

\begin{figure}[t!!]
\includegraphics*[keepaspectratio=true, clip=true, angle=0, width=1\columnwidth, trim={-35mm, 2mm, 50mm, 2mm}]{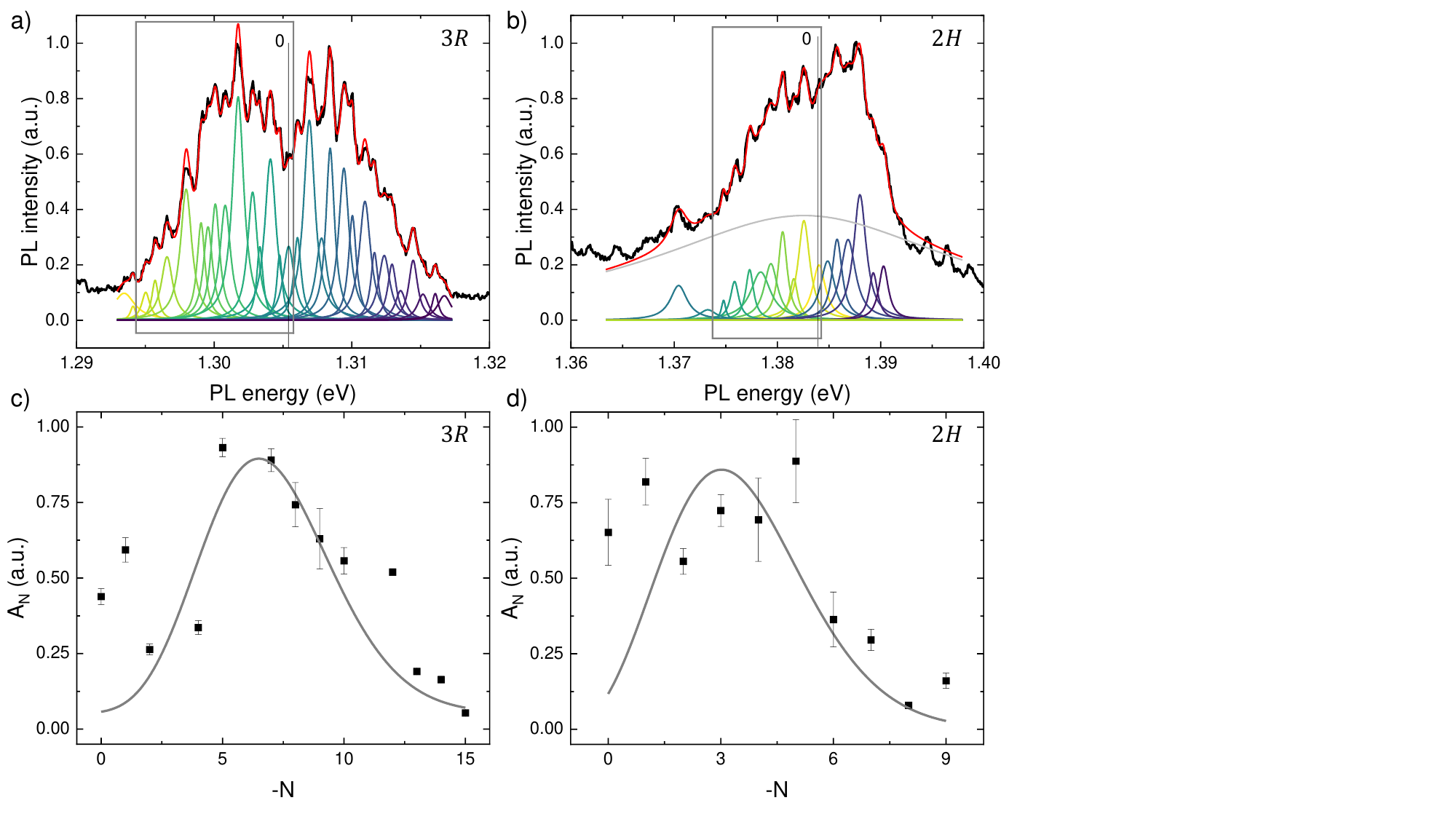}
\caption{\textbf{a)} and \textbf{b)} PL spectra for the 3$R$ and 2$H$ samples, respectively. On each spectra, the fitting of Lorentzian curves is presented in a colour scheme varying from yellow to violet. The experiments are in black and the fitting is superimposed in red. The grey curve in the 2$H$ sample correspond to the background. \textbf{c)} and \textbf{d)} Extracted intensities of the peaks marked with a rectangle in figure a and b. In grey, is the Poisson distribution fitting that results in a Huang-Rhys factor of $3.1\pm0.4$ and $1.5\pm0.2$ for the 3$R-$ and the 2$H-$sample, respectively. Note that in the 3$R-$sample, we did not consider the intensity of the peaks -3 and -6 because they presented an unexpected intensity. far above the rest of the distribution.}
\label{figura:SM-4}
\end{figure}

\subsection{Additional information regarding the temperature series}\label{Temperature Dep}

\begin{figure}[t!!]
\includegraphics*[keepaspectratio=true, clip=true, angle=0, width=1\columnwidth, trim={-70mm, 57mm, 290, 2mm}]{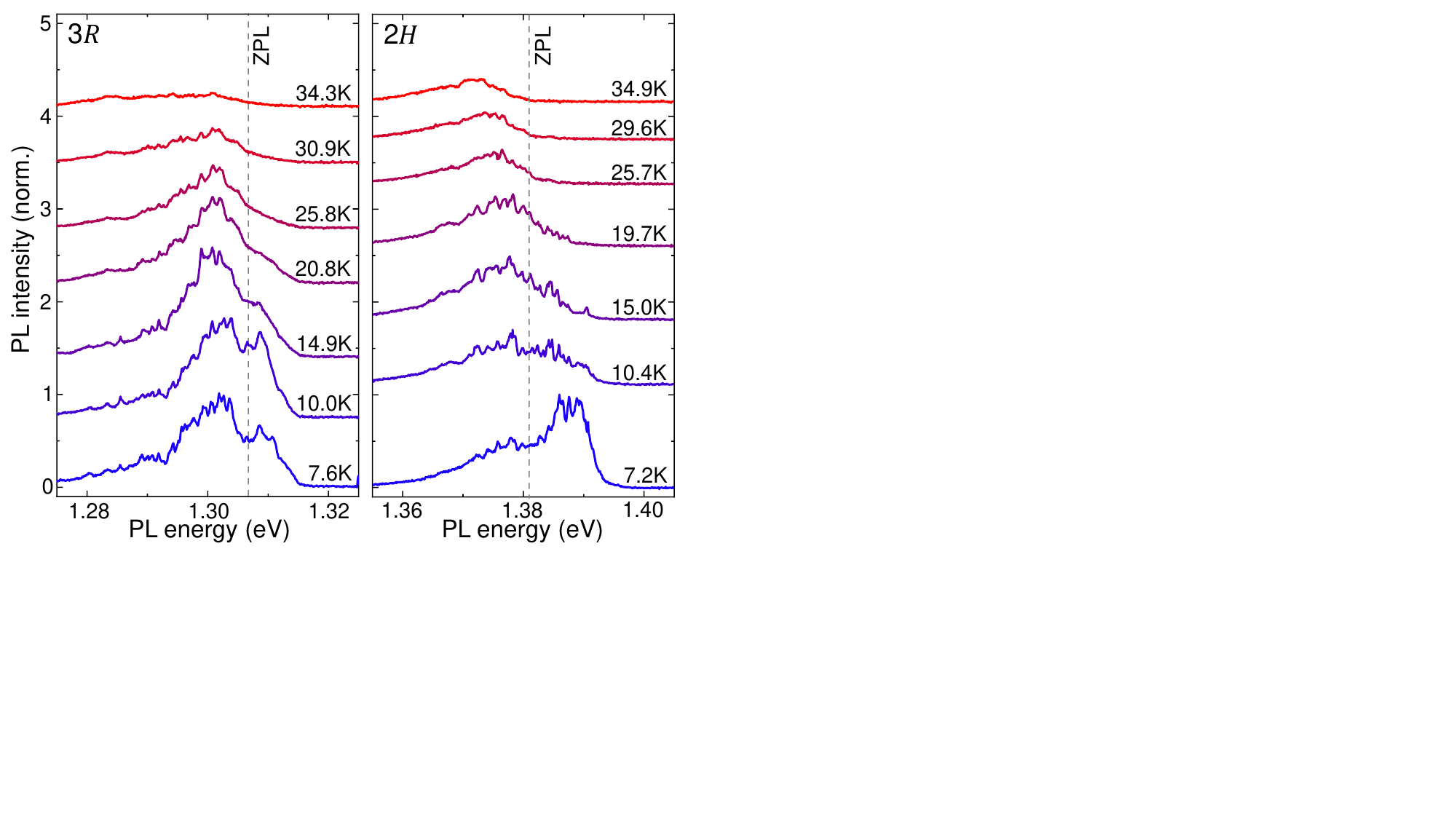}
\caption{Temperature dependent PL spectra for the 3$R-$sample (left) and the 2$H-$sample (right).}
\label{figura:SM6}
\end{figure}

In this section we complement the information regarding the temperature dependent PL presented in the main text by adding the data corresponding to the 3$R-$sample. The experiments were performed with a CW laser, $P_{ex} = 300$\,nW and $E_{exc}=1.96$\,eV. The spectra, normalized to their maximum intensity at 7\,K, are presented in Figure \ref{figura:SM6}, the left(right) panel present the data for the 3$R-$(2$H-$)sample. The grey dotted lines mark the position of the ZPL. Both samples display the same behaviour; at low temperature high and low energy sidebands show similar intensity but, by increasing the temperature, the high energy sideband is strongly suppressed. The narrow emission lines progressively fade out by increasing the temperature and, above $20$\,K for the high energy sideband and $30$\,K for the lower one, they are almost indistinguishable.

\subsection{Additional information regarding the power series}\label{Power Dep}

\subsubsection{Experiments}

\begin{figure}[t!!]
\includegraphics*[keepaspectratio=true, clip=true, angle=0, width=1\columnwidth, trim={-10mm, 57mm, 210, 2mm}]{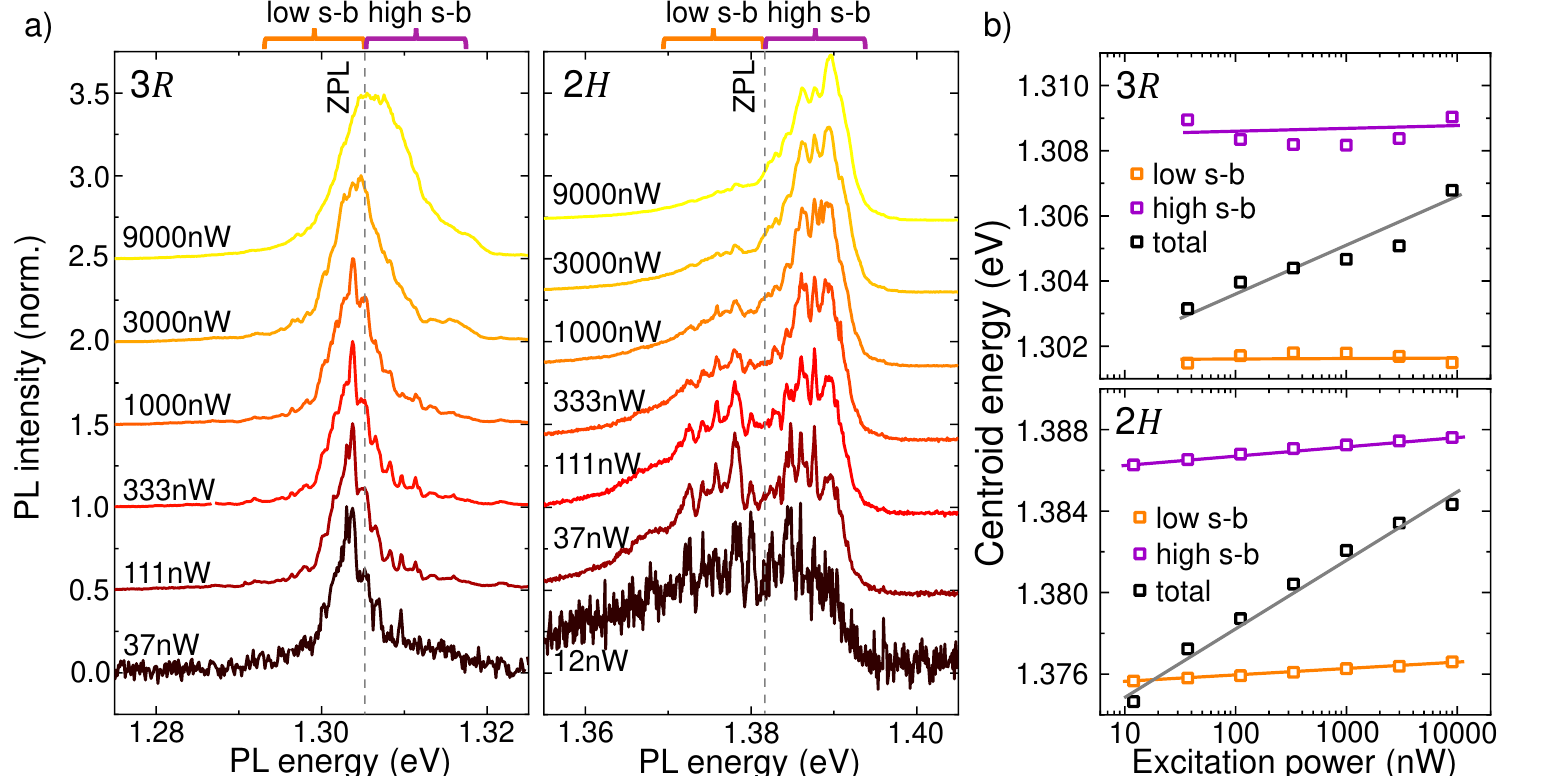}
\caption{\textbf{a)} Power dependent PL spectra for the 3$R-$sample (left) and the 2$H-$sample (right). On top of each panel, the region of $\pm12\,$meV from the ZPL is marked with orange(violet) to define the low(high) energy sideband. \textbf{b)} Centroid energy of the total spectra (black) and the low and high energy sideband in orange and violet, respectively.}
\label{figura:SM7}
\end{figure}

In this section we complement the information regarding the power dependence experiments presented in the main text by adding the data corresponding to the sample stacked at 4.5$^\circ$. In both samples, the experiments were performed at 7\,K, with a CW laser and $E_{exc}=1.96$\,eV. Figure \ref{figura:SM7}a present the power dependent PL, where left(right) panel correspond to the 3$R-$(2$H-$)sample. In both cases, the distinct emission peaks coalesce into a broad emission line by increasing $P_{ex}$. Additionally, there is a clear blueshift of the overall emission by increasing $P_{ex}$, in agreement with previous reports.

A careful observation of Figure \ref{figura:SM7}a reveals that the observed blueshift is due to a variation in the relative intensities between the low and high energy sidebands. Figure \ref{figura:SM7}b present the centroid energy of the spectra in Fig. \ref{figura:SM7}a the upper(lower) panel correspond to the 3$R-$(2$H-$)sample. Black square dots show, in both cases, the blueshift of the total IX emission centroid. This observation is completely different when we consider the centroid of each sideband, presented in orange(violet) dots for the lower(higher) energy sideband. In this case, it is clear that each sideband is spectrally fixed, and the $P_{ex}$ blueshift is currently an intensity increase of the upper energy sideband respect to the lower energy one. 

\subsubsection{Phenomenological description}\label{model}

\begin{figure}[t!!]
\includegraphics*[keepaspectratio=true, clip=true, angle=0, width=1.\columnwidth, trim={-20mm, 90mm, 200, 2mm}]{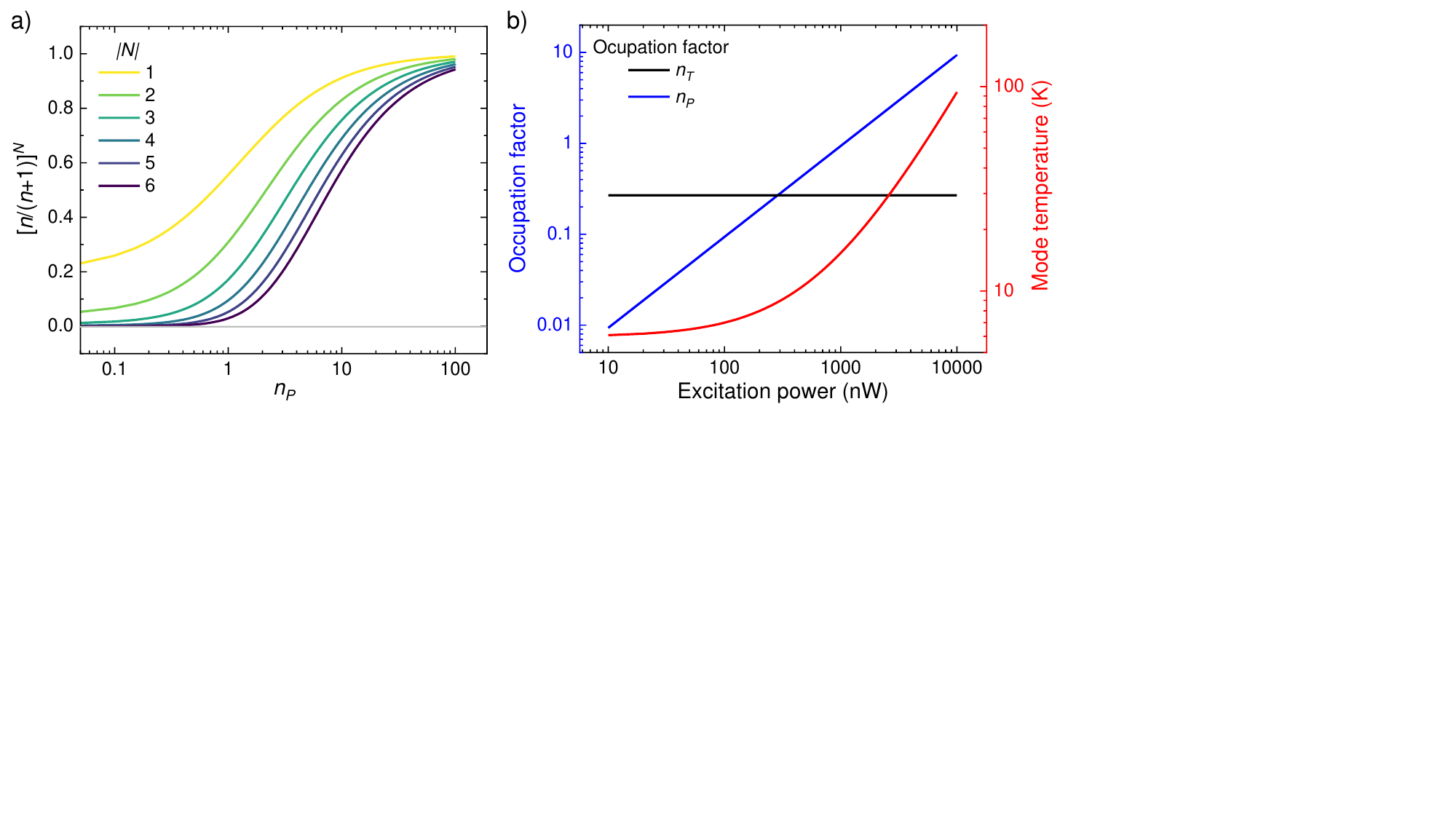}
\caption{\textbf{a)} $\frac{\tau_{N<0}}{\tau_{N>0}}\left(\frac{n_T+n_P}{n_T+n_P+1}\right)^N$ considering $n_T=0.25$ and $n_P$ varying from 0 to 100 for some $|N|$ varying from 1 to 6. \textbf{b)} Thermal and power dependent occupation factors and mode temperature derived from the $n_P$ and proportionality constant derived in the main text.
}
\label{figura:SM-8}
\end{figure}

We described the power dependence PL through a phenomenological model that accounts phonon replicas of arbitrary order under the following assumptions: i) The phonons that contribute to the sidebands are generated during the charge transfer and energy relaxation processes. Consequently, the phonon occupation number is $n = n_T+n_P$, were $n_T$ is the thermal occupation and $n_P$ are the optically generated phonons, ii) The $N^{th}$ emission line corresponds to $N-$ phonon absorption (for $N>0$) and $N-$ phonon emission (for $N<0$). Note that, strictly speaking, the $N^{th}$ emission line fulfills the following condition: $N=\alpha-\beta$, where $\alpha$($\beta$) corresponds to phonon absorption(emission) processes~\cite{feldtmann2009phonon} and, therefore, it is not possible to fully isolate absorption and emission processes. However, this approximation allows us to determine the relative intensity between the low and high energy sideband. It is well established in rate equation models that the number of radiative decays from a particular excited level and then its emission intensity in the spectra are proportional to $1/\tau$, where $\tau$ is the radiative lifetime of the transition. On the other hand, as the peaks we describes are phonon replicas of the ZPL, their abundance is a function of the phonon occupation factor. Following the peak labeling of the main texts, $N>0$ are IX emission in which the IX is dressed with $N$ phonons and its intensity in PL experiments is $n^{N}/\tau_N$. In the case $N<0$, however, there is always the possibility of an spontaneous emission and, therefore, the intensity of a phonon emission replica is $(n+1)^{-N}/\tau_N$. The ratio between positive and negative phonon replicas of the same order is then
\begin{equation}
\label{eq:model1}
    \frac{I_{N>0}}{I_{N<0}} = \frac{\tau_{N<0}}{\tau_{N>0}}\left(\frac{n}{n+1}\right)^N = \frac{\tau_{N<0}}{\tau_{N>0}}\left(\frac{n_T+n_P}{n_T+n_P+1}\right)^N.
\end{equation}
Consequently, the limit $n_P\rightarrow 0$ becomes 
\begin{equation}
\label{eq:model2}
    \lim_{n_P \to 0} \frac{I_{N>0}}{I_{N<0}} = \frac{\tau_{N<0}}{\tau_{N>0}}\left(\frac{n_T}{n_T+1}\right)^N
\end{equation}
and the limit $n_P\rightarrow \infty$
\begin{equation}
\label{eq:model3}
    \lim_{n_P \to \infty} \frac{I_{N>0}}{I_{N<0}} = \frac{\tau_{N<0}}{\tau_{N>0}}.
\end{equation}
Figure \ref{figura:SM-8}a shows for some $N$s the function $\frac{\tau_{N<0}}{\tau_{N>0}}\left(\frac{n_T+n_P}{n_T+n_P+1}\right)^N$ considering $n_T=0.25$ and $n_P$ varying from 0 to 100.

In the main text, we did not make any distinction between modes of different order and only considered the difference between high and low energy sideband. As result, we obtained an average behaviour of the different phonon replicas that, as expressed by eq. \ref{eq:model3}, allowed us to estimate the ratio between the low and high energy sideband lifetime. On the other hand, the low excitation power limit depends on $N$ (see eq. \ref{eq:model2}) and, therefore, the thermal population derived from it is underestimated.

Figure \ref{figura:SM-8}b presents, considering the parameters obtained in the main text, i.e. the average behaviour of eq. \ref{eq:model1}, the comparison between the thermal and power dependent occupation factor. The right axes displays the mode temperature, showing that at a relative low $P_{exc}$ the mode temperature is much higher than that of the lattice ($\sim7$\,K).

\twocolumngrid

%
%

\bibliographystyle{apsrev4-2}
\bibliography{bibliography}

\end{document}